\documentstyle[preprint,prl,aps]{revtex}

\newcommand{\be}{\begin{equation}}
\newcommand{\ee}{\end{equation}}
\newcommand{\BE}{\begin{eqnarray}}
\newcommand{\EE}{\end{eqnarray}}
\newcommand{\BEn}{\begin{eqnarray*}}
\newcommand{\EEn}{\end{eqnarray*}}
\newcommand{\barr}{\begin{array}} 
\newcommand{\earr}{\end{array}}
\newcommand{\bit}{\begin{itemize}}      
\newcommand{\eit}{\end{itemize}}
\newcommand{\bfl}{\begin{flusleft}}
\newcommand{\efl}{\end{flusleft}}
\newcommand{\bfr}{\begin{flushright}}
\newcommand{\efr}{\end{flushright}}

\newcommand{\bc}{\begin{center}}
\newcommand{\ec}{\end{center}}

\newcommand{\ben}{\begin{enumerate}}    
\newcommand{\een}{\end{enumerate}}

\newcommand{\eps}{\varepsilon}

\newcommand{\bs}{{\bf s}}
\newcommand{\bz}{{\bf 0}}
\newcommand{\bj}{{\bf j}}

\begin{document}
\draft
\tightenlines

%\sloppy

\title{N-tree approximation for the Largest Lyapunov Exponent of a Coupled 
map lattice}
\author{F.Cecconi$^{1,2}$ and A.Politi$^{2,3}$}
\address{  
$(1)$ Dipartimento di Fisica, Universit\`a di Firenze \\
$(2)$ INFN Sezione di Firenze\\
$(3)$ Istituto Nazionale di Ottica, Firenze, Italy}
\date{}

\maketitle
\begin{abstract}
The $n$-tree approximation scheme, introduced in the context of random
directed polymers, is here applied to the computation of the maximum
Lyapunov exponent in a coupled map lattice. We discuss both an exact
implementation for small tree-depth $n$ and a numerical implementation for
larger $n$s. We find that the phase-transition predicted by the mean field
approach shifts towards larger values of the coupling parameter when the
depth $n$ is increased. We conjecture that the transition eventually
disappears.

\end{abstract}
\vskip 1.cm

%\begin{multicols}{2}

\newpage
\narrowtext

%========================================================================
\section{Introduction}
%========================================================================

The development of analytical techniques to determine Lyapunov exponents
in extended systems is certainly an important issue in view of the
relevant information provided by them. One cannot, in general,
expect to find exact solutions, as the problem is already non-trivial
in the case of low-dimensional systems. Since most of the results published
in the literature about Lyapunov exponents follow from numerical simulations,
the development of effective ``perturbative'' techniques is very welcome
as they can also provide information about interactions and correlations 
that are otherwise undetectable.

More numerical results are certainly available for coupled-map lattices
(CML), since the discreteness of both space and time variables allow simpler 
and faster simulations. The commonest coupling scheme for a CML is
\BE
x^{t+1}_i &=& f (y^t_i) \nonumber \\
y^{t}_i &=& \eps x^t_{i-1} + (1-2\eps) x^t_i + \eps x^t_{i+1} 
\label{eq:cml}
\EE
The corresponding evolution equation in the tangent space is
\be
\xi^{t+1}_i = f'(y^t_i) 
\bigg(\eps \xi^t_{i-1} + (1-2\eps) \xi^t_i + \eps \xi^t_{i+1} \bigg) \quad .
\label{eq:cmlt}
\ee
from which one can see that even the computation of the maximum Lyapunov 
exponent (MLE) in a CML
requires the simultaneous consideration of several issues: (i) space-time
correlations of the local multipliers $m_i^t = f'(y^t_i)$; (ii) sign
fluctuations of the multipliers which induce partial cancellations in the
dynamics of the perturbation $\xi_i^t$; (iii) correlations in tangent space
induced by the spatial coupling.

The third of the above issues is definitely the first to be clarified
as it arises already in the presence of positive $\delta$-correlated
multipliers. It is precisely this problem that we shall address in the
present paper, trying to determine the MLE in the random matrix
approximation, i.e. assuming that all multipliers are independent,
identically distributed, random processes. This is the standard assumption
made in the study of Anderson localization in disordered systems so that we
can say that our investigation can be naturally extended to such a case.

A first attempt to determine the effect of spatial coupling has been
made in \cite{LPR}, where the authors performed a mean field anlaysis,
exploiting the analogy with the free-energy computation in directed polymers
growing in random media. There, it has been found that the spatial coupling
induces a shift in the value of the MLE from the quenched average ($\lambda
= \langle \log m_i \rangle$) in the absence of coupling, to the annealed
average ($\lambda = \log \langle  m_i  \rangle$) above a critical coupling
value.

In this paper we apply the so-called ``$n$-tree'' approximation scheme
\cite{DS88,CD89}, to obtain more refined analytical estimates of the MLE and
to test the convergence properties for increasing depth of the tree.
The growing evidence that many features of CML dynamics are indeed present
also in chains of oscillators and in partial differential equations suggests
that techniques developed for CML can be extended to such systems.

The paper is arranged as follows. In section II we recall the essential lines
of the $n$-tree approximation in directed-polymer theory and reformulate the
approach in the present context. Section III is devoted to the numerical
implementation, while the small coupling limit is investigated in
section~IV. Finally, in section V we present some remarks about the
problem of estimating the MLE in a coupled map lattice and recall the open
problems.

%========================================================================
\section{The method}
%========================================================================

In this section, we first recall the $n$-tree approximation in the context
of directed polymers, with reference to a (1+1)d structure. The approach is
then explicitely formulated for the determination of the Lyapunov exponent.

Let us consider all directed walks in a square lattice composed of the
following displacements, $\bj \to \bj + \bs$, where $\bj = (i,n)$ represents
a generic site, while $\bs \in \{ (-1,1), (0,1), (+1,1) \}$ (see, e.g., 
Fig.~1), and attribute
a random energy $e(\bj)$ to each site. The statistical problem amounts to
computing the partition function $Z_L(0)$ of all $L$-step walks departing
from the origin,
\be
  Z_L(0) = \sum_w \exp(-\beta E_w)  \quad ,
\ee
where the sum runs over all $3^L$ paths, $E_w$ is the sum of the
energies in the sites visited during the walk, and $\beta$ is the inverse
temperature.

One can write a recursive relation for $Z_L$,
\be
Z_{L+1}(\bz) = \exp(-\beta e(\bz) ) \sum_{\bs} Z_L(\bs)
\label{recur1}
\ee
with the initial condition $Z_0(\bz) = 1$. The free energy per unit length
is nothing but the exponential growth rate of $Z$ with $L$,
$F = - \lim_{L\to \infty}\langle \log Z_L \rangle /(L\beta)$, where
$\langle \cdot \rangle$ represents the average over independent disorder
realizations.

The main difficulty preventing an exact solution of the above problem is the
correlation among the partition functions appearing in the r.h.s. of
Eq.~(\ref{recur1}). The $n$-tree approximation scheme consists in iterating
$n$ times such a recursive relation (which automatically accounts for
all correlations up to $n$ steps),
\be
   Z_{L+n}(0) = \sum_{ \bs_1,\bs_2,\dots,\bs_n }
   \exp \left [-\beta \sum_i e({\bf S}_i) \right ] Z_L({\bf S}_n) \quad ,
\ee
where ${\bf S}_i = \sum_{m=0,i-1} \bs_m$, and in neglecting the
remaining  correlations, i.e. among the terms in the r.h.s. of the
above expression.

The problem can be solved exactly by introducing the generating function
\be
   H_L(x) = \left\langle \exp\{-\hbox{e}^{-\beta x}Z_L\} \right\rangle
\ee
which satisfies the recursive equation
\be
   H_{L+n}(x) = \left\langle \prod_{j=-n}^{n} H_{L}(x - {1\over\beta}\log M_j) 
   \right\rangle  \quad ,
\label{hl}
\ee
where $M_j$ is the contribution of all paths of length $n$ arriving at
site $j$ (see Fig.~1). The initial condition 
$H_0 \equiv \langle \exp \{- \hbox{e}^{-\beta x}\}\rangle$ is a front 
interpolating the two fixed points $P_0$ ($H=0$), and $P_1$ ($H=1$) of
Eq.~(\ref{hl}).
Since $P_1$ is unstable, while $P_0$ is stable, the front moves to the right. 
It can be easily seen that the front velocity is nothing but the free energy
of the polymer.

The velocity can be determined by approximating the forefront
as $H_L(x) \sim 1 - \exp(-\gamma x)$. Substitution of this Ansatz into
Eq.~(\ref{hl}) reveals that the front moves with a velocity that depends on
$\gamma$,
\be 
   c(\gamma) = \left\{\barr{ll} G_n(\gamma)     &  \gamma \le \gamma_{min}  \\
                              G_n(\gamma_{min}) &  \gamma > \gamma_{min}
                      \earr
               \right. \quad ,
\ee
where
\be
   G_n(\gamma) = \frac{1}{n\gamma} \log \left \{\sum_{j=-n}^{n}
   \langle M_j^{\gamma/\beta} \rangle \right \}
\ee
and ~$\gamma_{min}$~ is the $\gamma$-value where $G_n$ takes its minimum
\be
    \frac{dG_n(\gamma)}{d\gamma} \bigg\vert_{\gamma_{min}} = 0 \quad .
\ee
The value of $\gamma$ is implicitely determined by the initial condition:
by expanding $H_0(x)$ for large $x$, one realizes that $\gamma = \beta$.
Therefore, if $\beta < \gamma_{min}$, the free energy coincides with
the annealed average of the weights $M_j$, while different $\gamma$-averages
are selected at lower temperatures. This implies the existence of a
thermodynamic transition occurring at $\beta_c = \gamma_{min}$.

The analogy between the evolution equation (\ref{eq:cmlt}) in tangent space,
and the recursive relation (\ref{recur1}) for the partition function
suggests that the whole procedure can be extended to the computation
of the MLE. In this new context, the time $t$ plays the role of the
polymer length $L$ and $m_i^t$ can be considered as a quenched noise in a 2d
environment. The growth rate of $Z$ (i.e. the free energy), finally becomes
the Lyapunov exponent.

The differences with respect to the previous cases are the presence of
the anisotropy factors $\eps$ and $1-2\eps$ and the absence of a temperature. 
The first one is, in principle, only a technical variation which leads, 
however, to a strong ``degeneracy'' as we will show in the following. 
The temperature, instead, can be
removed by setting $\beta = 1$; the role of the relevant control parameter
will be played by the coupling strength $\eps$ parameter. 
With the above simple indications, we find that the $n$-tree approximation 
$\Lambda_n$ of the Lyapunov exponent is
\be
\Lambda_n   
   = \left\{ \barr{lc} G_n(\gamma_{min}) &  \quad \gamma_{min} < 1     \\
                       G_n(\gamma=1)     &  \quad \gamma_{min} \geq 1 
              \earr
     \right. \quad .
\label{eq:mle_rm}
\ee
The function $G_n(\gamma)$  has the form
\be
G_n(\gamma) = \frac{1}{n\gamma} 
\log \left\{ \sum_{j} \left \langle \left(
\sum_{w_j} (1-2\eps)^{k(w_j)} \eps^{n - k(w_j)} \prod_i m_i(w_j)
\right)^{\gamma} \right \rangle \right\} \quad .
\label{eq:genne}
\ee
As shown in Fig.~1, $w_j$'s are the directed walks on the lattice arriving
at site $j$ after $n$ steps (the depth of the tree), $k(w_j)$ is the number
of steps not involving a change of position, and $m_i(w_j)$ is the
multiplier in the $i$th time step of the path $w_j$.

Eq.~(\ref{eq:mle_rm}) suggests the possible existence a phase transition
upon changing $\eps$: if $\gamma_{min} \geq 1$, the Lyapunov exponent is
given by the annealed average of the multipliers.
The main difficulty in the implementation of this approach is the
computation of $G_n$ and its minimization. In the limit ~$n\to\infty$, the
approximation becomes exact: an interesting question concerns the
convergence to the asymptotic value. Unfortunately, as $n$ increases, the
expression of ~$G_n$~ quickly becomes so complicated that it is practically
impossible to handle the analytical expression. For this reason, in the next
section we shall address the question from the numerical point of view.

%========================================================================
\section{Numerical results}
%========================================================================

It is not only true that the analytical expression of the MLE becomes
very complicated as $n$ increases, but also an ``exact'' numerical
implementation is not an easy task. Indeed, the average of disorder implies 
the computation of several
multiple integrals and, even in the simple case of a uniform distribution
of multipliers, the presence of a power $\gamma$ in Eq.~(\ref{eq:genne})) 
makes the integrals immediately undoable. The 
only case we have found, where it is possible to combine an exact solution 
of the integrals with a powerful numerical analysis, is that of a dichotomic
distribution. Indeed, a generic multiple integral over $K$ variables becomes 
a sum over all $2^K$ combinations of the variables, each properly weighted
according to the probability of the two possible values of the multiplier.
From Fig.~1, one can see that the number $K$ of integrals to be performed is
already equal to 7 for for $n=3$ and $j=-1$, which in turn requires summing
up 128 different terms. As a result, even in this simple case, it is not
possible to go beyond $n=5$.

Since a global rescaling of the multipliers yields a trivial shift of
the MLE, we can assume, without loss of generality, that $m_i(t) =
\{1,b\}$. Moreover, for the sake of simplicity, and in order to maximize
the effect of the fluctuations (which are responsible for the deviation
of the MLE from the single-map case) we have assumed that 1 and $b$ have
the same probability 1/2.

The results for $n=1,\ldots,5$, $b=3$ and small $\eps$ values are reported in
Fig.~2, where a slow convergence towards the asymptotic value (numerically 
determined by iterating a chain of 1000 maps) can be observed. 
In the inset of the same figure, one can also notice that the critical
$\eps$-value, above which the Lyapunov exponent corresponds to the annealed
average, steadily increases, in agreement with the numerical results
that do not give any evidence for the existence of the high-temperature phase.

Although an exact implementation of the $n$-tree scheme is unfeasible
already for $n > 5$, one can consider it as a numerical algorithm. Indeed,
one can imagine to iterate $n$ times a perturbation initially localized in
the origin. The amplitude in the site $j$ represents an instance of $M_j$
and the average required in Eq.~(\ref{eq:genne}) can be computed by summing
over independent realizations of the stochastic process. 

Moreover, one can notice that $\gamma$ plays a similar role to $q$ in the 
standard multifractal analysis; the only difference is that here, besides a 
local $\gamma$-average, a linear average over different sites is also 
required. It is therefore important to understand how $\gamma_{min}$ behaves
for increasing $n$, i.e. to clarify whether the actual value of the MLE
does arise from a specific $\gamma$ value.

The only drawback of the numerical implementation is the need of a sufficient
statistics, a constraint that becomes increasingly important for larger
$n$-values, since an accurate determination of $M_j^\gamma$ strongly
depends on unprobable large deviations as usual in a multifractal analysis.
Notwithstanding this limitation, it has been possible to arrive at $n=50$,
much beyond the limits for an exact implementation.

The results for $\eps=0.01$ (reported in Fig.~3) confirm that the
approximate values of the MLE approach from above the asymptotic value 
$\Lambda_\infty$ (denoted by the horizontal line in the figure). A numerical 
investigation of
the behaviour of $(\Lambda_n - \Lambda_\infty)$ versus $n$ suggests that the
convergence is presumably slower than algebraic. The slow variation of
$\Lambda_n$ with $n$ is confirmed by the poor improvement obtained by
introducing the refined estimates
\be
\tilde \Lambda_n = {n \Lambda_n - (n-5)\Lambda_{n-5}\over 5} \quad ,
\label{eq:refined}
\ee
(see the squares in Fig.~3), where the choice of ``5'' is simply dictated by
the spacing of the numerical results. Notice that this procedure is very
effective when the main finite-size effect arises from some rapidly
decaying initial deviation, as it is the case of the Lyapunov exponent
computed with the standard orthonormalization procedure \cite{galga}.

Besides allowing to compute the MLE, the $n$-tree approximation yields an 
estimate of the optimal $\gamma_{min}$ value. By comparing the results
for the various depths, one finds that $\gamma_{min}$ slowly
decreases. We conjecture that $\gamma_{min}$ eventually converges to 0. The
conclusion is suggested by the analogy between Eq.~(\ref{eq:genne}) and the
behaviour of the maximal comoving Lyapunov exponent $\lambda_c(v)$ \cite{kane},
which is defined as the growth rate in the site $i = vt$ of a perturbation
initially localized in the origin. In a system with left-right symmetry,
$\lambda_c(v)$ reaches its maximal value for $v=0$, where it coincides with
the MLE. Therefore, for $n$ large enough, the dominant contribution to
$G_n(\gamma)$ is given by the growth rates around the origin, i.e. by
their logarithmic average. Accordingly, we expect that $\gamma_{min}$ will
eventually approach 0.

Such a tendency is confirmed for all $\eps$-values that we have considered,
even well inside the supposed high-temperature phase where $\Lambda_n =
G_n(1)$. One such example is illustrated in Fig.~4, where we report
$G_n(\gamma)$ for $\eps = 0.05$ and different depths $n$. Notice that all
$G_n(\gamma)$-curves take the same value for $\gamma=1$; this is a
consequence of the very definition of $G_n(\gamma)$: independently of $n$,
$G_n(1)$ coincides with the annealed average. For relatively small values of
$n$, the minimum of $G_n$ is attained for $\gamma_{min} > 1$ and the correct
estimate of the Lyapunov exponent is given by the annealed average. However,
upon increasing $n$, $\gamma_{min}$ steadily decreases until $\gamma_{min}<1$
(in the case illustrated in Fig.~4, this happens for $20<n<30$). The
conjectured convergence of $\gamma_{min}$ to zero implies the eventual
disappearance of the phase-transition.

Further light onto the $n$-tree approximation scheme can be shed by
comparing it with a similar, though entirely heuristic, approach.
The structure of $G_n(\gamma)$ requires iterating an initially localized
perturbation: it is therefore quite natural to consider a different initial
condition, uniformly spreaded over $2n$ sites, and to iterate it for $n$
steps by assuming periodic boundary conditions. At variance with the scheme
understood in Eq.~(\ref{eq:genne}), all sites are now statistically
equivalent, so that we can estimate the MLE directly from their growth rate
(or, better, from the average growth rate, to reduce statistical
fluctuations). The performance of this empirical approach can be judged from
the results reported in Fig.~3, where one can see that the convergence is
now from below. Moreover, while the direct estimates are worse than the
corresponding values obtained from the $n$-tree scheme, the opposite is
true for the improved estimates. We can interpret this results as an
indication that the actual value of the MLE in an extended system follows
from a delicate balance of several processes. A more effective reduction
of finite-size effects can be presumably accomplished by introducing a
still different definition of finite-time finite-space Lyapunov exponent.
Whether one such a definition exists which applies to generic models is not
obvious at all.

%========================================================================
\section{Coupling sensitivity}
%========================================================================

One case that is worth investigating with the aid of the $n$-tree
approximation is the small-coupling limit. In particular, it is interesting
to study the scaling behaviour of the MLE for decreasing $\eps$.
This problem has been already considered in Ref.\cite{LPR}, with the help of
a mean field approach. Here, we discuss the improvements arising from the
implementation of the $n$-tree scheme.

In this section, $\eps$ will be always so small that even in the lowest
approximation ($n=1$), $\gamma_{min} <1$ and the MLE is given by the
low-temperature expression. In the $1$-tree approximation, the Lyapunov
exponent is given by $\Lambda(\eps) = G_1(\gamma_{min}(\eps))$, where
$G_1$ follows from Eq.~(\ref{eq:genne}) with $n=1$,
\be
G_1(\gamma) = {1\over{\gamma}} \log\langle m^{\gamma} \rangle + 
 {1\over\gamma} \log\bigg\{ (1-2\eps)^{\gamma} + 2\eps^{\gamma} \bigg\}
\quad .
\label{eq:G1}
\ee
For small $\eps$, the above expression simplifies to
\be
G_1(\gamma) \cong {1\over\gamma}
\log \langle m^{\gamma} \rangle +
 {1\over\gamma} \log (1 + 2 \eps^{\gamma}) \quad .
\ee
Since, for $\eps\to 0$, one must recover the value of the uncoupled case,
~$\gamma_{min}(\eps)$~ must go to zero in such a way that also
$\eps^{\gamma_{min}(\eps)} \to 0$. Accordingly, one can further expand
Eq.~(\ref{eq:G1}), obtaining
\be
G_1(\gamma) \cong  \langle \log m \rangle + {\Gamma_2\over 2} \gamma +
2\;{\eps^{\gamma}\over\gamma}
\ee
where $\Gamma_2 = \langle (\log m)^2 \rangle - \langle \log m \rangle^2$ is
the variance of the local expansion rate. With some algebra, it is possible
to show that the value of $\gamma_{mim}$ minimizing ~$G_1$~ is given by
\be
\gamma_{mim}(\eps) = 2\; \frac{\log\sigma}{\sigma} \quad ,
\label{eq:gamm1}
\ee
where $\sigma \equiv |\log\eps|$ diverges for $\eps \to 0 $. On the one
hand, formula (\ref{eq:gamm1}) confirms the correctness of the Ansatz
$\gamma_{min}(\eps)\to 0$ and that $\eps^{\gamma}=1/\sigma$ can be really
considered a small parameter. On the other hand, we notice that the
convergence to zero is extremely slow.

By substituting (\ref{eq:gamm1}) into the expression for $G_1$ and
retaining only the leading terms in ~$\eps$, we obtain
\be
\Lambda_1(\sigma) = \Lambda_0  +  \Gamma_2 \frac{\log\sigma}{\sigma}
+ \frac{1}{\sigma\log\sigma} \quad ,
\label{eq:lyap1}
\ee
where $\Lambda_0 = \langle \log m \rangle$ is the Lyapunov exponent of the
uncoupled problem, the second term is responsible for the leading
correction and the third smaller contribution is reported for the sake
of completeness.

The above result, already derived in \cite{LPR}, has been here recalled
because it allows introducing all the key steps that are necessary to deal
with higher-order approximations. In the following, we illustrate the case
$n=2$ with some detail and mention the further adjustments expected for
larger depths of the tree. The complete expression for
$G_n(\gamma)$ is already rather complicated for $n=2$,
\be
G_2(\gamma) = {1\over {2\gamma}} \log \left\{f_1(\gamma)\right\}
+ {1\over 2\gamma} \log \bigg\{ 2\eps^{2\gamma}
f_1(\gamma) + 2\eps^{\gamma}(1-2\eps)^{\gamma} f_2(\gamma) + f_3({\gamma})
\bigg\} \quad ,
\label{eq:G2}
\ee
where
\BEn
 f_1(\gamma)&=&\langle m^{\gamma}\rangle  \\
 f_2(\gamma)&=&\langle ( m_1 + m_2 )^{\gamma}\rangle \\ 
 f_3(\gamma)&=&
 \langle (\eps^{2} m_1 + (1-2\eps)^{2} m_2 + \eps^{2} m_3)^{\gamma} \rangle 
\EEn
while $m$, $m_1,m_2,m_3$ represent the multipliers to be averaged.

With the aid of the same approximations made for ~$n=1$, the above
expression simplifies to
\be
G_2(\gamma) \cong \langle \log m \rangle + {\Gamma_2 \over2}\gamma + 
{\eps^{\gamma}\over{\gamma}}
\ee
which, in turn,  gives the following expression for the Lyapunov exponent
\be
\Lambda_2(\sigma) = \Lambda_0 +  \Gamma_2 \frac{\log\sigma}{\sigma}
+ \frac{1}{2}\; \frac{1}{\sigma\log\sigma} \quad .
\label{eq:lyap2}
\ee
By comparing Eqs.~(\ref{eq:lyap2}) and (\ref{eq:lyap1}), one can see that
they differ only in the last term of the r.h.s., which is smaller by a factor
two in the 2-tree approximation. The same is true (apart from the
coefficient of the third term) for larger values of $n$. Accordingly, this
analysis seems to indicate that the MLE grows as
$\log|\log \eps| / \log \eps$, independently of the depth $n$. These
conclusions are indeed confirmed by the numerical implementation of the
$n$-tree scheme for several depths. This can be noticed in Fig.~5, where
it is reported the behaviour of
\be
  \Sigma_n  \equiv ( \Lambda_n - \Lambda_0 ) |\log \eps|
\label{sn}
\ee
versus $\log \eps$. The slow growth of $\Sigma_n$ is
consistent with the expected $\log | \log \eps |$ behaviour. However,
direct numerical simulations (full dots in Fig.~5) suggest that the MLE
grows as $1/\log \eps$, with no doubly logarithmic correction. In our
opinion, the apparent
contradiction can be solved by noticing that the determination of the
correct scaling behaviour requires, in principle, to take first the limit
$n \to \infty$ (to estimate correctly the MLE) and then the limit
$\eps \to 0$. In the above analysis, we have, instead, exchanged the two
limits. Since the determination of $G_n(\gamma)$ requires estimating an
exponentially growing (with $n$) number of contributions, it is reasonable
to conjecture that increasingly small $\eps$-values must be reached before
the leading term really overtakes the others. It is therefore conceavable 
that, before this asymptotic regime sets in, a different scaling region 
appears which becomes wider and wider for increasing $n$. The several limits 
involved in this process (we must not forget the role of $\gamma$) make a 
rigorous confirmation of this conjecture a rather delicate matter. Here, we 
limit ourselves to recall that in two coupled 1d and 2d maps 
a purely $1/\log \eps$ behaviour has been proved to arise \cite{daido}. 
However, we cannot exclude that in the present case the scaling $\eps$-region 
is so small that it is has not been reached by our numerical simulations; it 
will be possible to give a definite answer only by developing a more effective 
perturbative technique.

Finally, let us notice that the naive approach outlined in the second
part of the previous section falls short of identifying the leading
$1/|\log \eps|$ dependence, predicting only linear corrections in $\eps$.
We suspect that this failure is due to the lack of an infinite time limit
in the corresponding definition of the Lyapunov exponent.

%========================================================================
\section{Conclusions and perspectives}
%========================================================================

In this paper we have implemented the $n$-tree approximation for the
computation of the MLE. The results have revealed a slow convergence
towards the asymptotic value. We believe that more than being an intrinsic
limitation of the method, this is an indication of the hardness of the
problem. This feeling is indeed confirmed by the analogy with the
Kardar-Parisi-Zhang (KPZ) equation. As already shown in \cite{piko}, the
logarithm $h = \log \xi$ of the perturbation approximately satisfies the KPZ
equation
\be
h_t = \eps h_xx + \eps (h_x)^2 + \xi(t) \quad ,
\ee
where the subscripts denote derivatives with respect to either time ($t$) or
space ($x$) variables (assumed now to be continuous) and $\xi(t)$ is a
noise-like term corresponding to the logarithm of the local multiplier.
Accordingly, the average value of $\xi$ is nothing but the single-map
Lyapunov exponent, while the MLE of the lattice is obtained by adding the
average of the nonlinear term. Such a correction can be easily estimated by
recalling that in 1d, the probability distribution of $h$ is exactly the
same as for the linear Edwards-Wilkinson model, obtained by neglecting the
nonlinearity \cite{basta}. The latter is the distribution of the standard
Brownian motion, i.e. the product of independent Gaussians for the spatial
derivatives $h_x$ \cite{basta}. Since the variance of each Gaussian is
inversely proportional to the coefficient of the Laplacian, it is seen that
$\eps \langle h_x^2 \rangle$ is independent of $\eps$. Therefore, the KPZ
equation provides for any coupling strength the same annealed average value,
i.e. it even fails to find the low-temperature phase. This result indicates 
that one must go beyond the KPZ approximation of the tangent dynamics 
(adding higher-order derivatives and further nonlinear terms) if the 
$\eps$-dependence of the MLE is to be recovered.

The identification of effective schemes to fasten the convergence of
finite-size estimates of the MLE remains an open problem. Possible routes
to be explored in the future are represented by corrections to the KPZ
equation, or by suitable modifications of the $n$-tree approximation.

\vspace{1.0cm}
{\bf Acknowledgments} We thank the Institute of Scientific Interchange 
(I.S.I.) in Torino Italy, where this work was started.

%%%%%%%%%%%%%%%%%%%%%%%%%%%%%%%%%%%%%%%%%%%%%%%%%%%%%%%%%%%%%%%%%%%%%%%%

\begin{figure}
\caption{Schematic diagram of all paths starting from site 0 and
arriving at site $j$ in 3 steps}
\end{figure}

\begin{figure}
\caption{$n$-tree approximations of the MLE for $n=1,\ldots,5$ (from top to
bottom) compared with the numerical results (dashed curve). These and all
the other results in the paper have been obtained for a dichotomic
distribution of the local multipliers ($m=1,3$, with equal probabilities).
An enlargement of the region around the supposed phase-transition is
reported in the inset.}
\end{figure}      

\begin{figure}
\caption{Results of the $n$-tree approximation for $\eps=0.01$
(bullets) and refined estimates as from Eq.~(\ref{eq:refined}) (squares).
Triangles and diamonds refer to the results of the ``naive'' approximation
described in the end of section III and to the corresponding refinements,
respectively. The horizontal line represents the MLE as determined with
the standard numerical procedure.}
\end{figure}

\begin{figure}
\caption{$G_n$ vs. $\gamma$ for $n=20$ (dashed line), $n=30$
(dot-dashed line) and $n=40$ (full line) obtained from the numerical 
implementation of the $n$-tree approximation with $\eps = 0.05$.}
\end{figure}

\begin{figure}
\caption{ $\Sigma_n$ (see Eq.~(\ref{sn})) vs. $\eps$ for different values
of $n$. The solid lines correspond to $n=1,\ldots,5$ (from top to bottom).
Squares denote the results of the $n=40$ approximation, while bullets
correspond to the outcome of direct numerical simulations.}
\end{figure}
%\end{multicols}      

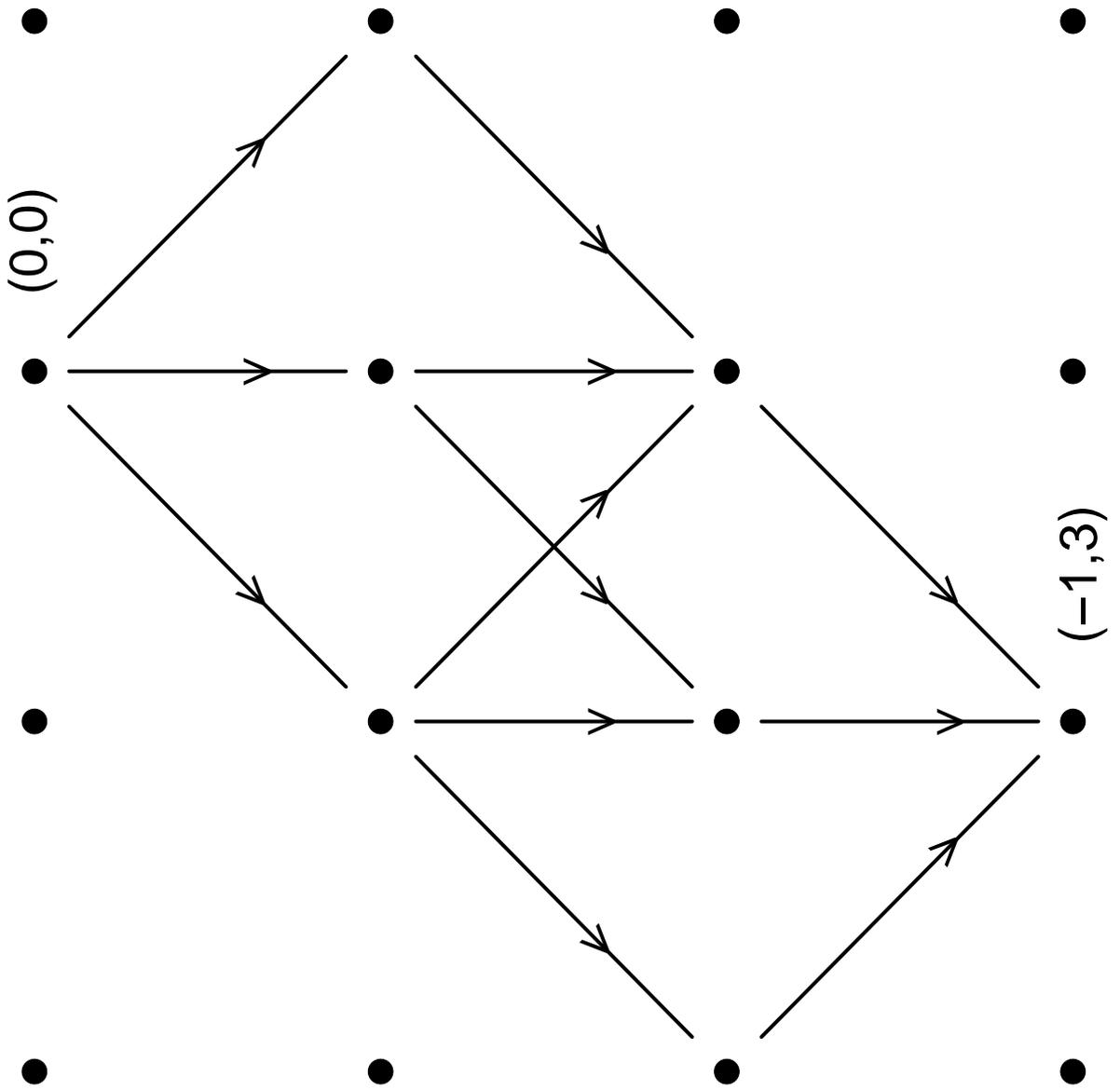
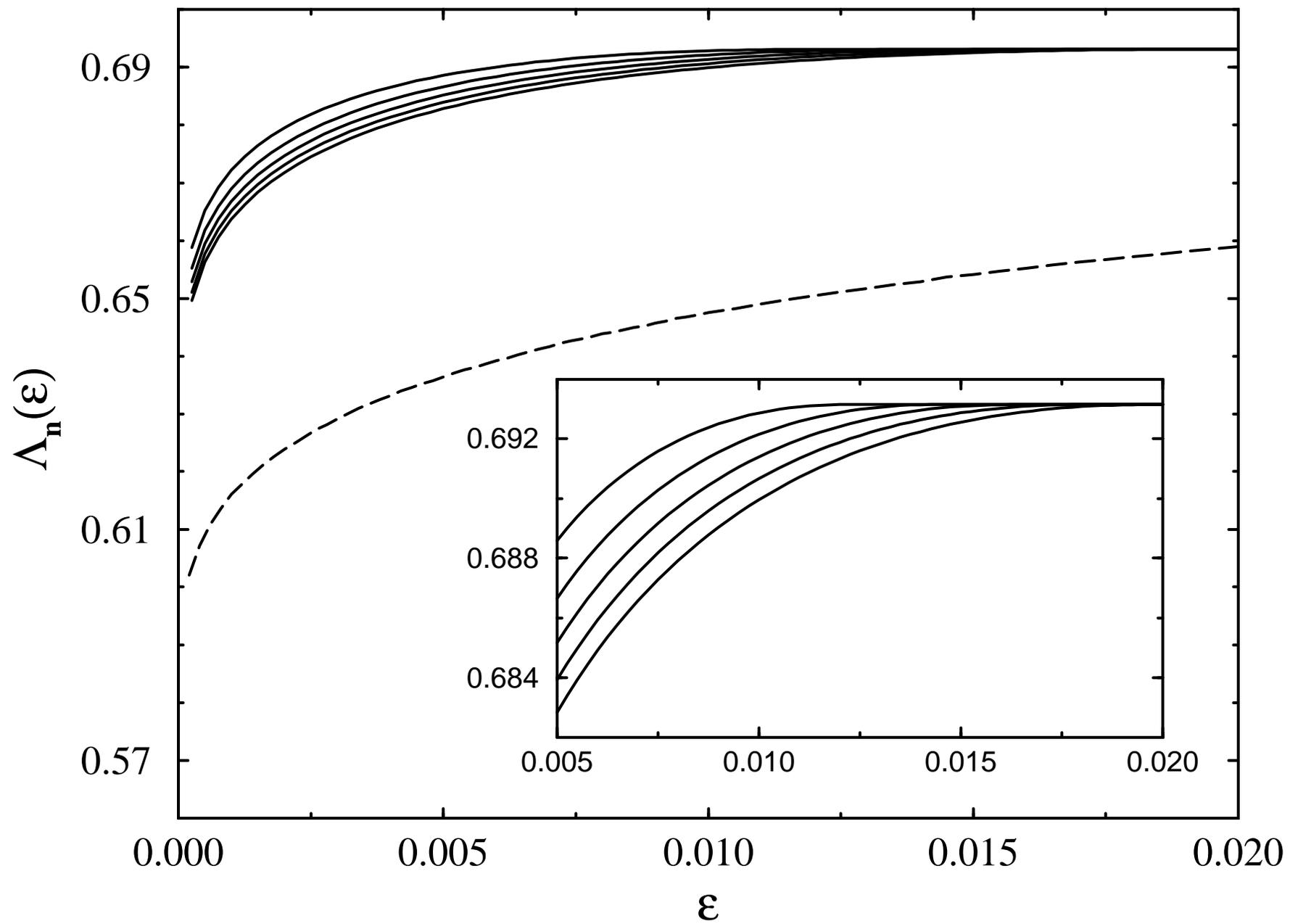
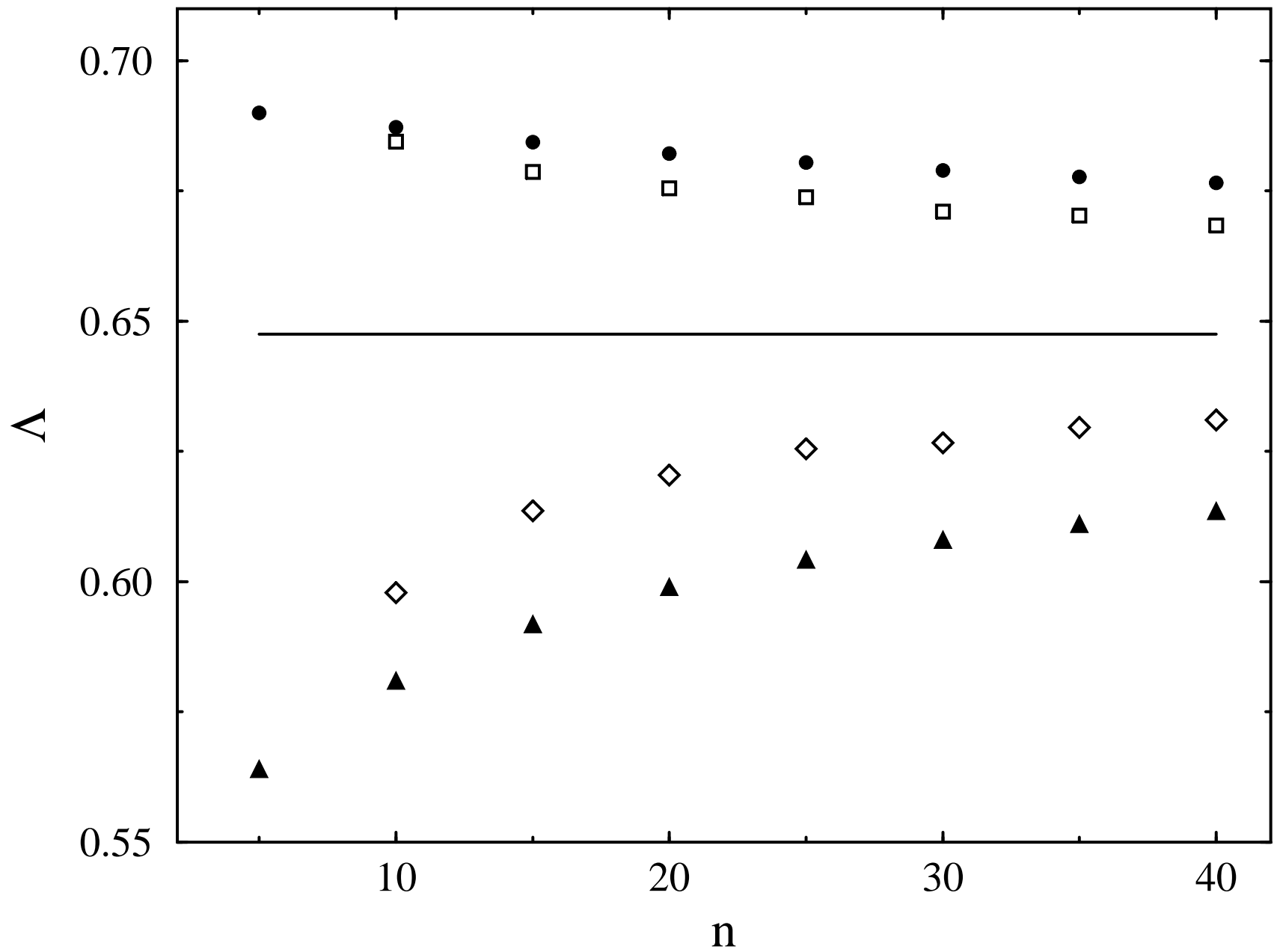
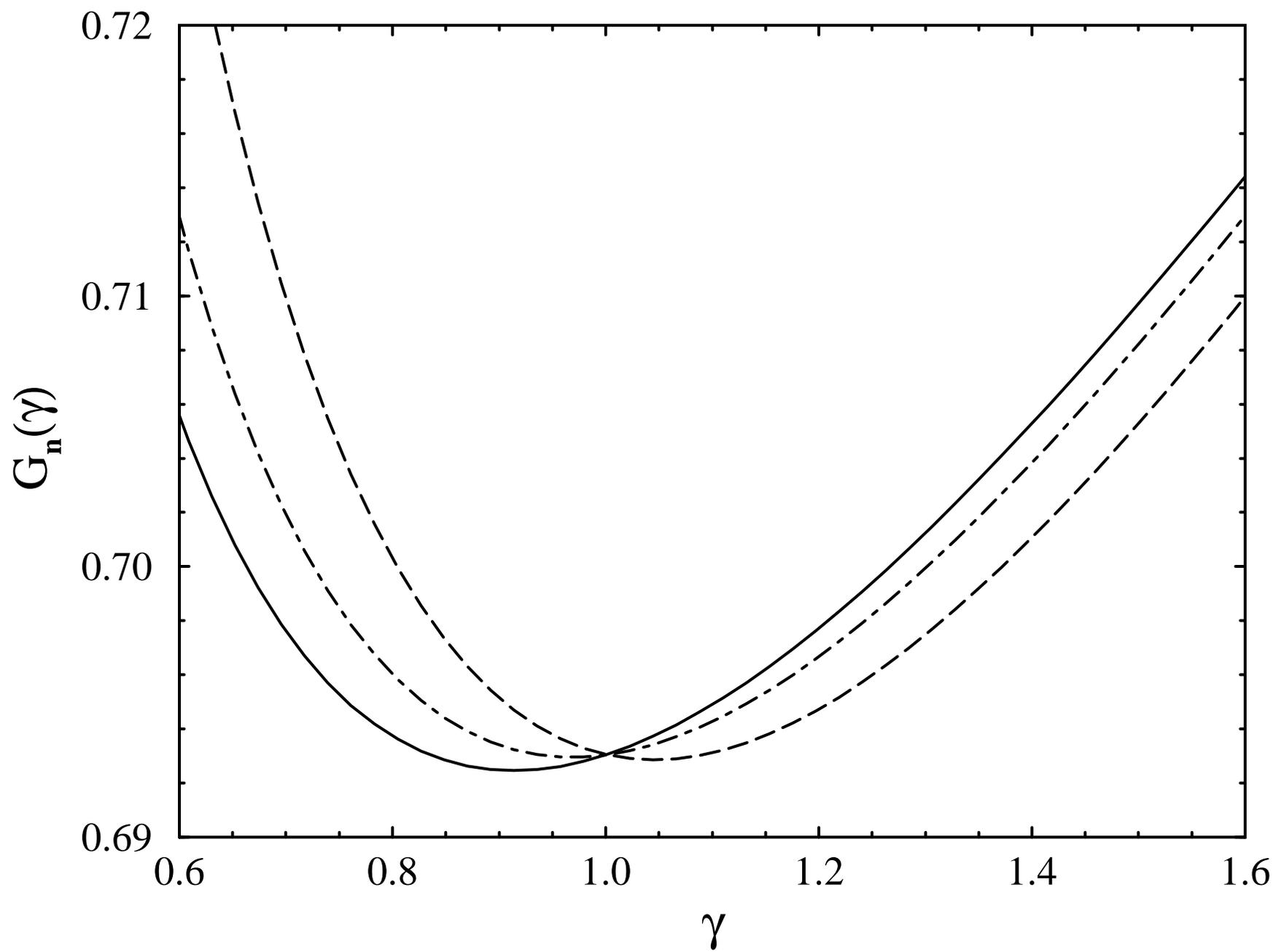
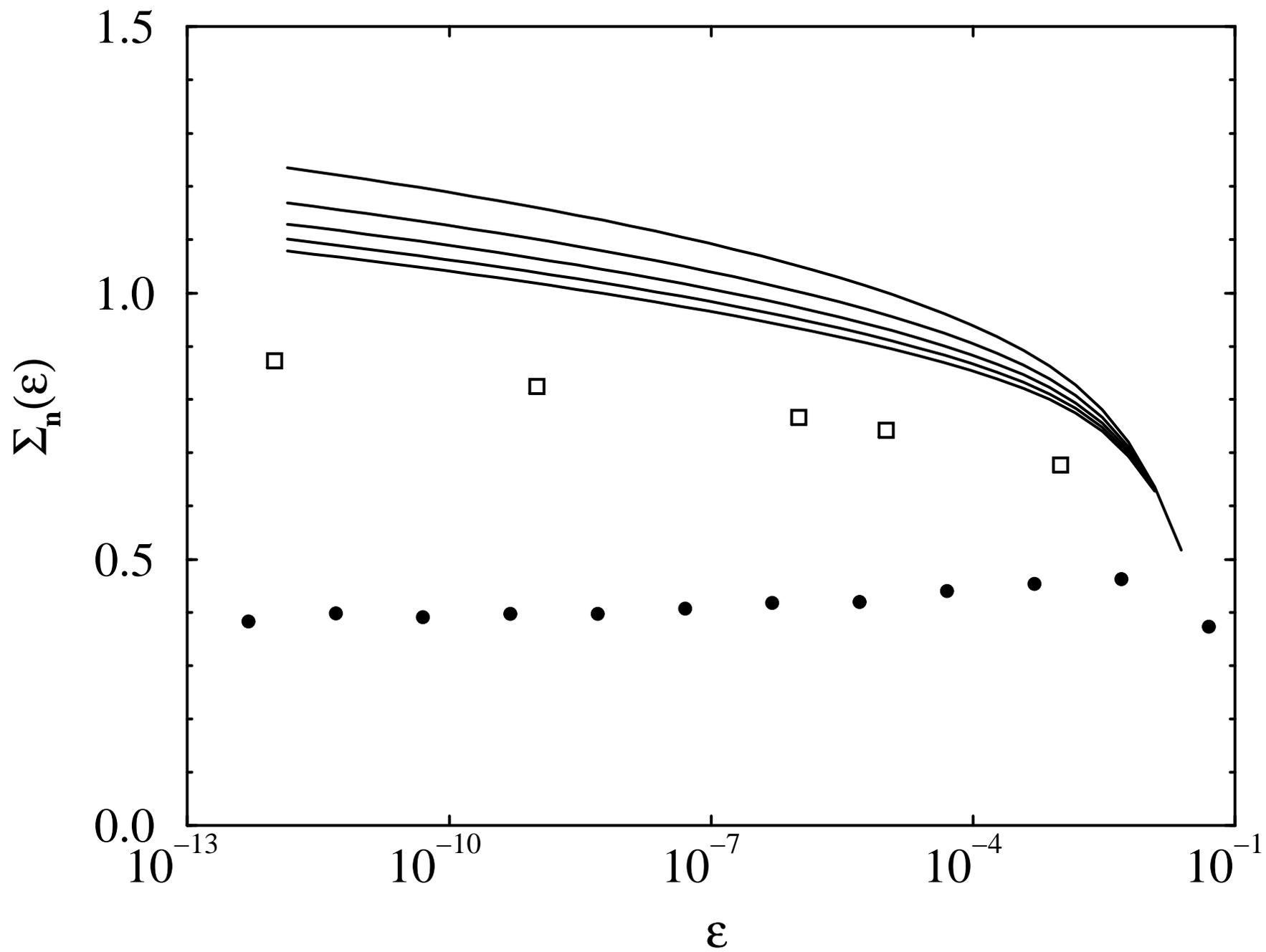
\end{document}